\documentclass[prl,twocolumn, showpacs, preprintnumbers,amsmath,amssymb,superscriptaddress]{revtex4-1}
\usepackage{graphicx}
\usepackage{dcolumn}
\usepackage{bm}
\usepackage{subfigure}
\usepackage{multirow}
\usepackage{amsmath}
\usepackage{color}
\usepackage[font=footnotesize, justification=raggedright,singlelinecheck=false]{caption}
\begin{document}

\title{Self-assembled multi-layer simple cubic photonic crystals of oppositely charged colloids in confinement}

\author{Krongtum Sankaewtong}
\thanks{These authors contributed equally.}
\affiliation{School of Chemical and Biomedical Engineering, Nanyang Technological University, 62 Nanyang Drive, 637459, Singapore}

\author{Qun-li Lei}
\thanks{These authors contributed equally.}
\affiliation{School of Chemical and Biomedical Engineering, Nanyang Technological University, 62 Nanyang Drive, 637459, Singapore}

\author{Ran Ni}
\email{r.ni@ntu.edu.sg}
\affiliation{School of Chemical and Biomedical Engineering, Nanyang Technological University, 62 Nanyang Drive, 637459, Singapore}

\begin{abstract}
Designing and fabricating self-assembled open colloidal crystals have become one major direction in soft matter community because of many promising applications associated with open colloidal crystals. However, most of the self-assembled crystals found in experiments are not open but close-packed. Here by using computer simulation, we systematically investigate the self-assembly of oppositely charged colloidal hard spheres confined between two parallel hard walls, and we find that the confinement can stabilize multi-layer NaCl-like (simple cubic) open crystals. The maximal layers of stable NaCl-like crystal increases with decreasing the inverse screening length. More interestingly, at finite low temperature, the large vibrational entropy can stabilize some multi-layer NaCl-like crystals against the most energetically favoured close-packed crystals. In the parameter range studied, we find upto 4-layer NaCl-like crystal to be stable in confinement. Our photonic calculation shows that the inverse 4-layer NaCl-like crystal can already reproduce the large photonic band gaps of the bulk simple cubic crystal, which open at low frequency range with low dielectric contrast. This suggests new possibilities of using confined colloidal systems to fabricate open crystalline materials with novel photonic properties.
\end{abstract}

\maketitle
\section{Introduction}
One of the major goals in colloidal self-assembly is to devise new colloidal systems to self-assemble into low-density open crystalline structures~\cite{torquato2009,cohn2009,edlund2011,chennature2011,romano2012,khalil2012,glotzer2007}, which are partially due to their promising applications in photonics~\cite{lopez2011}, catalysis, porous media~\cite{catabook} and the special response to mechanical stress~\cite{Souslov2009,kapko2009,sunpnas2012}. However, dated back to Maxwell, it has been proven that to maintain the mechanical stability, the coordination number of particles connected by central force bonds in periodic lattices must be larger than $2d$ in $d$ dimensions~\cite{maxwell}. Therefore, available methods for fabricating open colloidal crystals typically involve patterning the surface of colloidal particle forming anisotropic patchy particles with directional interactions~\cite{glotzer2005,glotzer2007,romano2012,chennature2011,romano2010jcp,wangyufengnature,glotzer2012,glotzer2013,glotzer2004}. It was also found that the vibrational entropy in the crystal formed by patchy particles stabilizes the low-density open crystal against the close-packed crystalline structures with the same energy~\cite{mao2013natmat,mao2013pre}.
However, making high quality monodisperse patchy colloids with well defined patches, which can self-assemble into open crystals, has been challenging in experiments~\cite{chennature2011,chenjacs2011,chenjacs2012,wangyufengnature,wangyufengjacs2013,gongnature2017,pinerev2013}. 

In past decades, a significant amount of scientific attention has been devoted into the self-assembly of colloidal systems in confinement, which has become a general way of tuning colloidal self-assembly~\cite{fortini2006phase,bart2015,wangnatcomm2018,glotzerpnas2016,lowenpre2018,lowenprl2012,lowenrev2009,lowenepl2009,lowen2005,Schmidt1997,Schmidt1996,peng2015,peng2015prl,glotzer2008}. However, most of colloidal crystals found in confinement are close-packed. Here by using computer simulation, we investigate the self-assembly of oppositely charged colloidal hard spheres confined between two parallel hard walls. Compared to the self-assembly of oppositely charged colloids in 3D systems~\cite{hynninen2006cuau,leunissen2005ionic}, we find the existence of stable multi-layer NaCl-like (simple cubic) crystals, of which the square plane is parallel to the confining walls. The maximal number of layers for stable NaCl-like crystal at the ground state increases with decreasing the inverse screening length, and the vibrational entropy at finite low temperature also help to stabilize the open multi-layer NaCl-like crystal against the energetically most favoured close-packed crystals.

\begin{figure}[b!]
\centering
\includegraphics[width=0.4\textwidth]{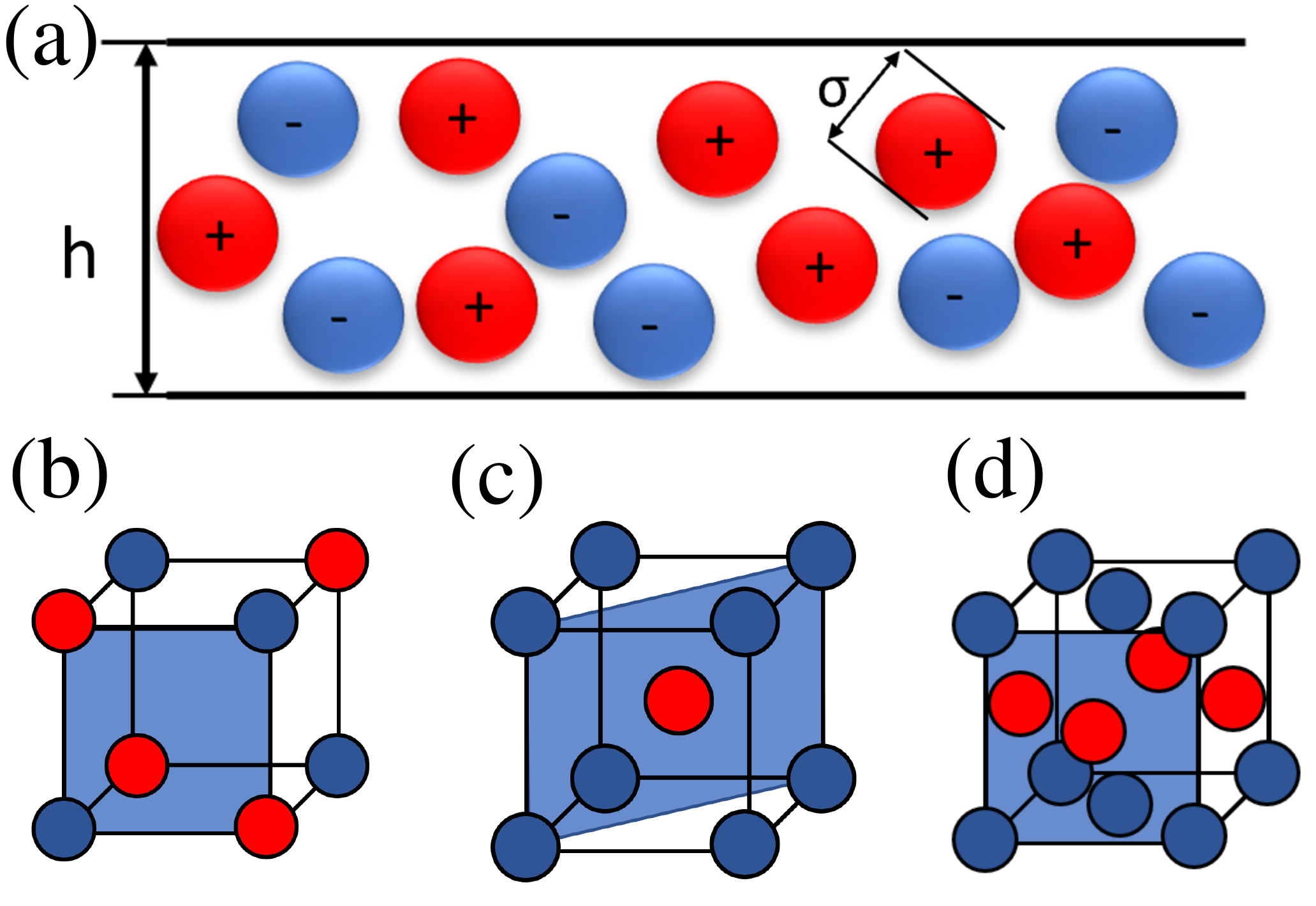}
\caption{(a) Illustration of oppositely charged particles (red or blue spheres) with diameter $\sigma$ confined between parallel hard walls with separation $h$. (b-d) Unit cells of ground states: NaCl (b), CsCl (c) and CuAu (d) crystals of the confined systems, where the shadow plane is parallel to the confining walls.}\label{fig1}
\end{figure}

\begin{figure*}[htbp]
\centering
\includegraphics[width=\textwidth]{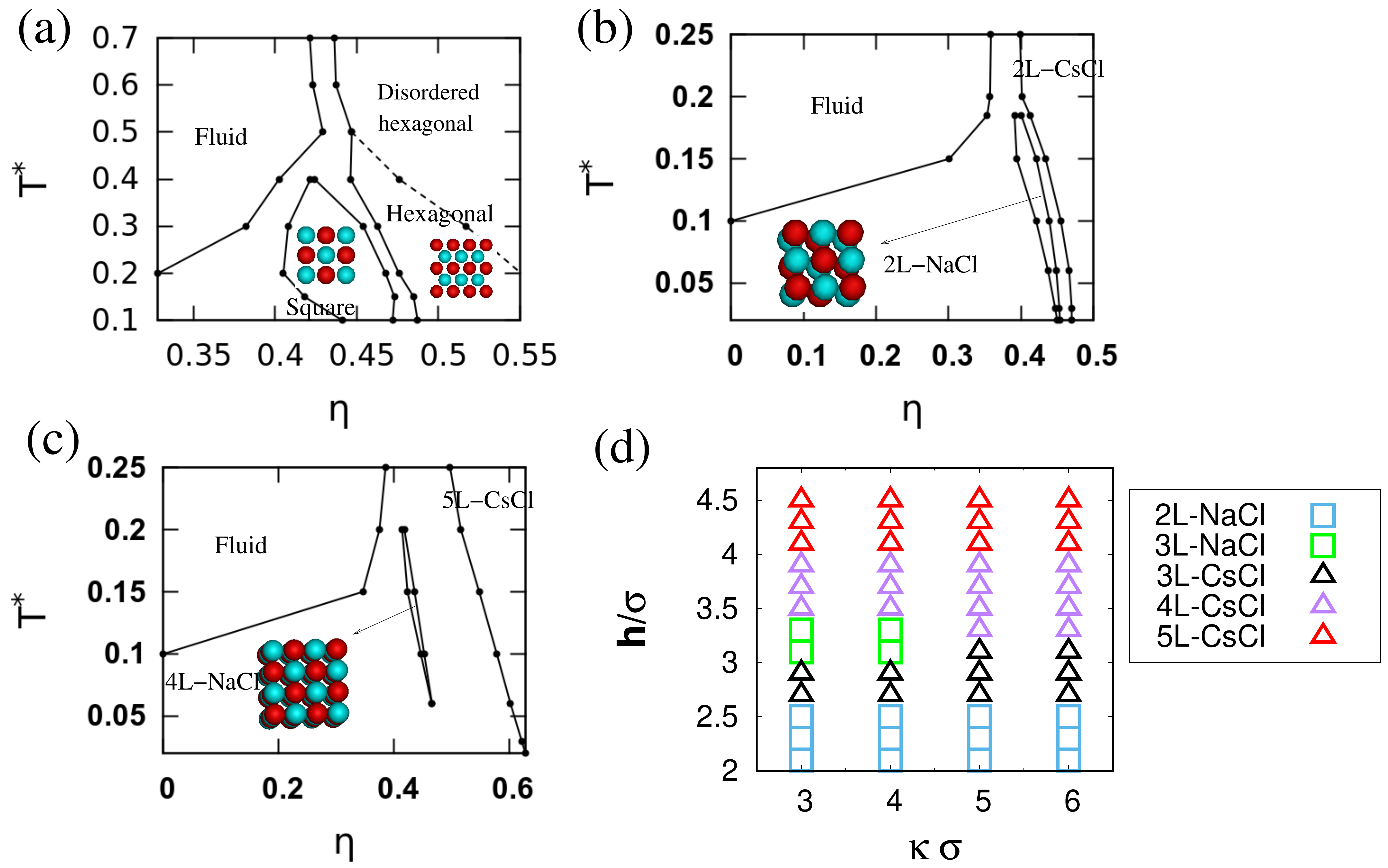}
\caption{Phase diagrams in the representation of $\eta$ versus $T^*$ for oppositely charged colloids confined between parallel hard walls with various separation $h/\sigma = 1$ (a), 2.3 (b), and 4.3 (c), where $\kappa = 3\sigma^{-1}$. {(d) The ground-state phase diagram for confined systems of oppositely charged colloids in the representation of $h/\sigma$ versus $\kappa \sigma$ at $T^* = 0$.}}\label{fig2}
\end{figure*}

\section{Methodology}
\subsection{Model}
We consider a system of $N$ colloidal spheres of a diameter $\sigma$ confined between two parallel hard walls in $z$ direction, half of which carry a positive charge $Ze$, and the other half a negative charge $-Ze$. The wall-wall separation is $h$, as shown in Fig.~\ref{fig1}a. We employ the Derjaguin-Landau-Verwey-Overbeek screened Coulomb pair potential~\cite{verwey1999theory} to model the interaction between charged colloids $i$ and $j$
\begin{equation}\label{eq1}
\frac{U(r_{ij})}{k_BT}= \begin{cases} 
      \pm \frac{Z^2}{(1+\kappa \sigma/2)^2} \frac{\lambda_B}{\sigma}      
      \frac{e^{-\kappa(r-\sigma)}}{r/\sigma} & r\geqslant\sigma \\
      \infty & r<\sigma 
   \end{cases},
\end{equation}
where $r_{ij}$ is the center-to-center distance between two particles, and $\lambda_B=e^2/\epsilon_sk_BT$ is the Bjerrum length with $\epsilon_s$ the dielectric constant of the solvent. 
$k_B$ and $T$ are the Boltzmann constant and the temperature of the solvent, respectively.
$\kappa = \sqrt{8 \pi \lambda_B \rho_{salt}}$ is the inverse Debye screening length with $\rho_{salt}$ the salt concentration. Here we focus on systems with $\kappa \ge 3\sigma^{-1}$, in which the pair potential in Eq.~\ref{eq1} was shown being a good approximation for modelling the screened Coulomb interaction in the system, and the multi-body effect is negligible~\cite{hynninen2004}. We define the packing fraction of the system as $\eta=\pi\sigma^3N/(6Ah)$ with $A$ the area of the system in $x-y$ plane, and the reduced temperature $T^*=(1+\kappa \sigma)^2\sigma/Z^2\lambda_B$, i.e., the inverse of the contact value of the potential in Eq.~\ref{eq1}. 
We assume that the confining walls are neutral and the interaction between particle $i$ and the confining walls is dominated by the excluded volume effect
\begin{equation}\label{eq2}
U_{wall}(r_i)=\begin{cases} 
0 & r_i\geqslant\sigma/2 \\
\infty & r_i<\sigma/2 \\
\end{cases},
\end{equation}
where $r_i$ is the distance of particle $i$ to the nearest wall in $z$ direction.

\subsection{Free energy calculation}
\subsubsection{Fluid phase}
We employ standard $NPT$ Monte Carlo simulation to obtain the equation of state (EOS) for the fluid phase, and we determine the free energy by integrating the EOS from the reference density $\rho_0$ to $\rho$:
\begin{equation}
\frac{F(\rho)}{N} = \frac{F(\rho_0)}{N} + \int_{\rho_0}^{\rho} \frac{P(\rho')}{\rho'^2} d \rho',
\end{equation}
where $F(\rho_0)/N = \mu(\rho_0) - P(\rho_0)/\rho_0$ is the Helmholtz free energy per particle at density $\rho_0$, and the chemical potential $\mu(\rho_0)$ is determined by the Widom's insertion method~\cite{frenkel2001}.

\subsubsection{Crystal phases in confinement}
For calculating the free energy of a crystal, we use the Einstein integration method~\cite{frenkel_ladd}. 
The Helmholtz free energy $F$ of a crystal confined between two parallel walls is
\begin{eqnarray}\label{eqint}
F(N,V,T) &=& F_{Einst}(N,V,T) \nonumber \\
&& -\int_{0}^{\lambda_{\max}}d\lambda \left \langle \frac{\partial U_{Einst}(\lambda)}{\partial \lambda} \right \rangle_{\gamma_{max},1/T^*_{\max}}  \nonumber \\
&& +\int^{1/T^*}_{0} d (1/T^*) \left \langle \frac{\partial \sum_{\substack{ i\neq j}}U(r_{ij})}{\partial (1/T^*)} \right \rangle_{\gamma_{max},\lambda_{max}} \nonumber \\
&& +\int^{\gamma_{\max}}_{0}d\gamma \left \langle \frac{\partial \left [ \sum_{\substack{i\neq j}} \varphi(i,j) + \sum_{i} \psi(i)\right]}{\partial \gamma} \right \rangle_{1/T^*=0,\lambda_{\max}},
\end{eqnarray}
where $F_{Einst}$ is the free energy of the ideal Einstein crystal. $\varphi(i,j)$ and $\psi(i)$ are the softness potentials between particle $i$ and $j$ and the nearest wall, respectively:
\begin{equation}
\frac{\varphi(i,j)}{k_BT} = \begin{cases} 
      \gamma[1-A(1+\zeta(i,j))] & \zeta(i,j) < 0 \\
      0 & otherwise \\
   \end{cases}
\end{equation}
and
\begin{equation}
\frac{\psi(i)}{k_BT} = \begin{cases} 
      \gamma[1-A(1+\zeta(i))] & \zeta(i) < 0 \\
      0 & otherwise \\
   \end{cases}
\end{equation}
where $\zeta(i,j)$ and $\zeta(i)$ are the surface distances between particle $i$ and $j$ and between particle $i$ and its nearest wall, respectively, and they are negative when overlapped. $\gamma$ is the integration parameter with $A=0.9$~\cite{fortini2006phase}. Here we choose to use $\gamma_{\max} = 150$ to guarantee the hard-core interaction is reproduced at $\gamma_{\max}$. $U_{Einst}(\lambda)$ is a harmonic potential to fix the particles onto a perfect crystal lattice:
\begin{equation}
U_{Einst}(\lambda) = \lambda \sum_{i}(\mathbf{r}_i-\mathbf{r}_{i,0})^2/\sigma^2,
\end{equation}
where $\mathbf{r}_i$ and $\mathbf{r}_{i,0}$ are the position of particle $i$ and its corresponding position in a perfect lattice, respectively. $\lambda$ is the integration parameter, and we choose $\lambda_{\max} = 2000$. 
At $h/\sigma = 1$, the crystal is exactly in 2D, and the free energy for the ideal Einstein crystal in 2D is~\cite{qi2013phase} 
\begin{equation}
\frac{F_{Einst}(N,V,T)}{N k_BT} = -\frac{N-1}{N}\ln \left( \frac{\pi}{ \lambda_{\max}}\right) - \frac{\ln A \sigma^2}{N}.
\end{equation}
For $h/\sigma > 1$, the reference ideal Einstein crystal is in 3D, of which the free energy is~\cite{ni2012phase}
\begin{equation}
\frac{F_{Einst}(N,V,T)}{N k_BT} = -\frac{3(N-1)}{2N}\ln \left( \frac{\pi}{ \lambda_{\max}}\right) + \frac{\ln \rho \sigma^3}{N} - \frac{3\ln N}{2N}.
\end{equation}
For calculating the integrations in Eq.~\ref{eqint}, we use the 20-point Gaussian Quadrature integration method. We have ensured that error in the calculated free energy is smaller than $0.01k_B T$ per particle.

\section{Results}
\subsection{Phase diagram}
We first investigate the phase behaviour of the system with $\kappa = 3\sigma^{-1}$ for different $h$. We perform floppy box Monte Carlo (MC) simulations with the potential in Eq.~\ref{eq1} and \ref{eq2} in isothermal-isobaric ensemble and periodic boundary conditions in $x$ and $y$ directions~\cite{laura2009prl}, in which the screened Coulomb potential is truncated at $r_{cut} = 3.5\sigma$. By quenching the simulation to $T^* \rightarrow 0$ at various $h$, we find three candidate ground states as shown in Fig.~\ref{fig1}b-d. When $T^* \rightarrow 0$, at low pressure, the system may form NaCl-like (simple cubic) structures confined between two walls, and with increasing pressure $P$, it switches to be confined CsCl-like (body-center-cubic) structures and CuAu-like (face-centered-cubic) structures at high density. With these candidate crystal structures, we employ the Einstein integration method~\cite{frenkel_ladd} combined with thermodynamic integrations to calculate the free energy of crystals in confinement~\cite{frenkel2001,fortini2006phase}. We compare the free energy of crystals with those of the fluid phase obtained from Widom insertion method~\cite{widom} to construct the phase diagrams of the system at $h/\sigma = 1, 2.3$ and 4.3 as shown in Fig.~\ref{fig2}a-c.

When $h/\sigma = 1$, the confined system is exactly in 2D. As shown in Fig.~\ref{fig2}a, at the low density limit, because of entropy, the stable phase is always a disordered fluid. With increasing the density, crystals form. At large $T^*$, the fluid crystallizes into a hexagonal crystal with disordered particle charges, and when decreasing $T^*$, a few charge-ordered crystals appear in the phase diagram. For $T^* \lesssim 0.4$, an ordered binary square crystal nucleates from the fluid at low pressure, and with increasing pressure, the high density charge-ordered hexagonal crystal emerges. The stability of the square crystal persists to $T^* \rightarrow 0$, as the square crystal is the energetically most favoured crystal of the system in 2D. Similar behaviour of the system at $h/\sigma = 2.3$ is shown in Fig.~\ref{fig2}b, in which a 2-layer NaCl-like crystal is more stable than the high density 2-layer CsCl-like crystal at $T^* \rightarrow 0$ because of the fact that the low density crystal is more energetically favoured. It is known that in 3D systems of oppositely charged colloids, CsCl is more stable than NaCl~\cite{hynninen2006cuau}, which suggests that at large enough $h$, the stable multi-layer NaCl should disappear.
Therefore, we calculate the phase diagram for the confined system of oppositely charged colloids with $h/\sigma = 4.3$ in Fig.~\ref{fig2}c. One can see that the stability of 4-layer NaCl crystal at $T^* \rightarrow 0$ disappears, while the stable co-existence of 4-layer NaCl crystal with low density fluid still exists in the low temperature range $0.05 \lesssim T^* \lesssim 0.2$.

\begin{figure}[htbp]
\centering
\includegraphics[width=0.5\textwidth]{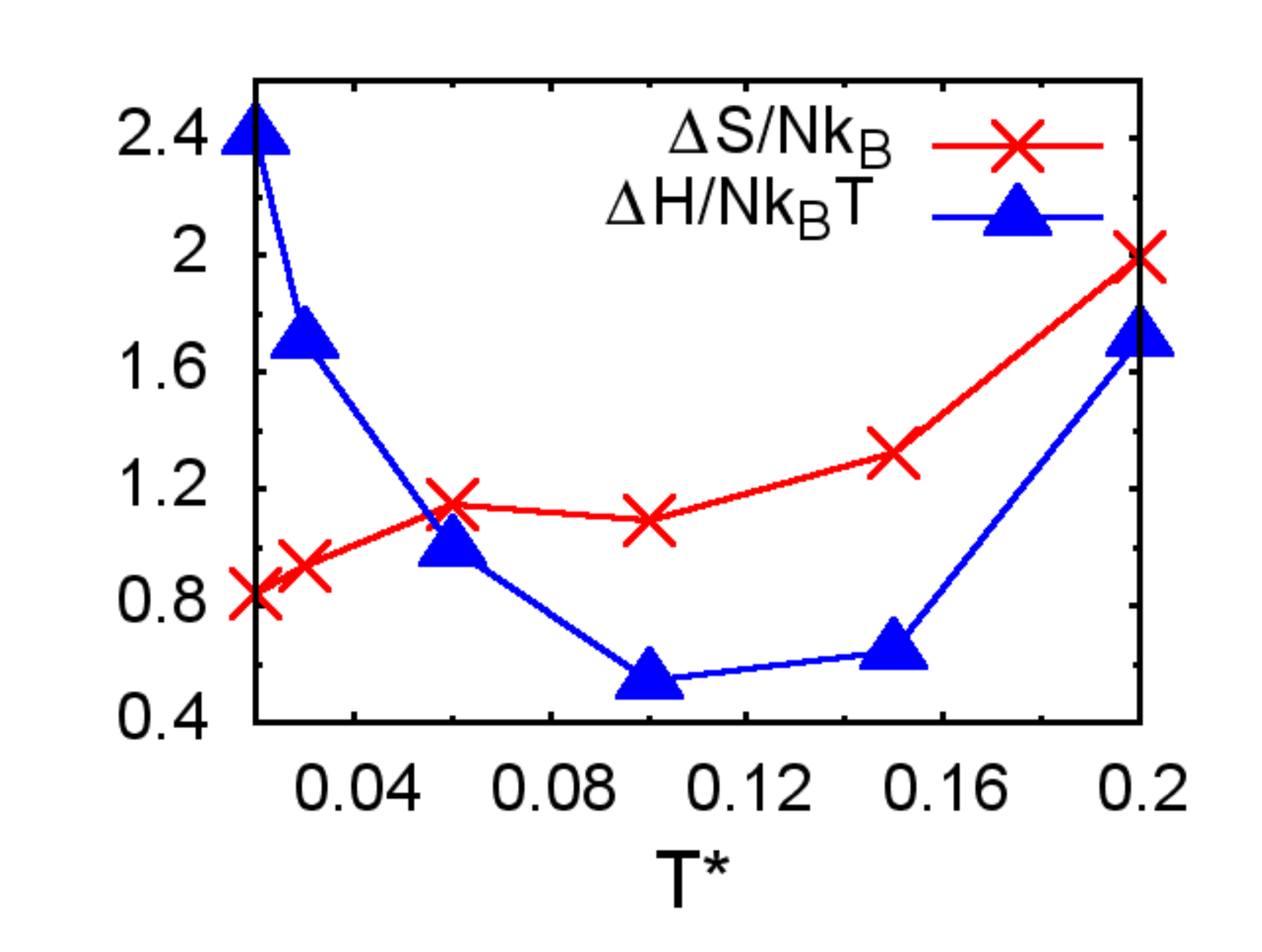}
\caption{The entropy difference $\Delta S/Nk_B = (S_{\mathrm{4L-NaCl}} - S_{\mathrm{5L-CsCl}})/Nk_B$ and the enthalpy difference $\Delta H /Nk_B T = (H_{\mathrm{4L-NaCl}} - H_{\mathrm{5L-CsCl}}) / Nk_B T $ as functions of $T^*$ at the pressure where the fluid becomes unstable in the confined oppositely charged colloids with $\kappa = 3\sigma^{-1}$ and $h = 4.3\sigma$. Here $S_{\mathrm{4L-NaCl} (\mathrm{5L-CsCl})}$ and $H_{\mathrm{4L-NaCl} (\mathrm{5L-CsCl})}$ are the entropy and enthalpy of 4-layer NaCl-like (5-layer CsCl-like) crystal, respectively.
}\label{fig4}
\end{figure}

In Fig.~\ref{fig2}d, we calculate the ground-state phase diagram for confined systems of oppositely charged colloids, and one can see that for $\kappa = 3\sigma^{-1}$, when $h/\sigma \gtrsim 3.5$, the multi-layer NaCl-like crystals becomes unstable with respect to the multi-layer CsCl-like crystals, and the maximal stable layers of NaCl-like crystals at the ground state is 3, which decreases with increasing $\kappa$. This agrees with our phase diagram in Fig.~\ref{fig2}c, in which the stable 4-layer NaCl-like crystal disappears at $T^* \rightarrow 0$. In Fig.~\ref{fig4}, to understand the stability of 4-layer NaCl-like crystal at low temperature, we plot the entropy $\Delta S /Nk_B$ and enthalpy $\Delta H/Nk_B T$ differences between 4-layer NaCl-like and 5-layer CsCl-like crystals in the confined system with $h/\sigma = 4.3$ at the pressure where the fluid becomes unstable for various $T^*$. Here the 5-layer CsCl-like crystal is the stable ground-state phase. One can see that for the entire range of $0 < T^*  \le 0.2$, $\Delta S$ and $\Delta H$ are always positive. This implies that vibrational entropy in the confined NaCl-like crystal is larger than that in the confined CsCl-like crystal, which is enthalpically more favoured.
Depending on the sign of the Gibbs free energy difference between the two crystals, i.e. $\Delta G = \Delta H - \Delta S T $, one can identify the stable phase.
At $T^* \rightarrow 0$, the enthalpy determines the stability $\Delta G > 0$, i.e. $\Delta H > \Delta S T$, and the first crystal nucleating from the low density fluid is the CsCl-like crystal. However, when $T^* \gtrsim 0.05$, $\Delta G < 0$, i.e. $\Delta H < \Delta S T $, which implies that the large vibrational entropy stabilizes the NaCl-like crystals against the enthalpically favoured CsCl-like crystal. This explains the stability of 4-layer NaCl-like crystal at finite low $T^*$ in Fig.~\ref{fig2}c, which is essentially an entropically stabilized low temperature solid phase of open crystalline structure. {This effect is similar to the entropy stabilized open crystals of patchy colloids~\cite{mao2013natmat,mao2013pre}.}

\begin{figure*}[htbp]
\centering
\includegraphics[width=\textwidth]{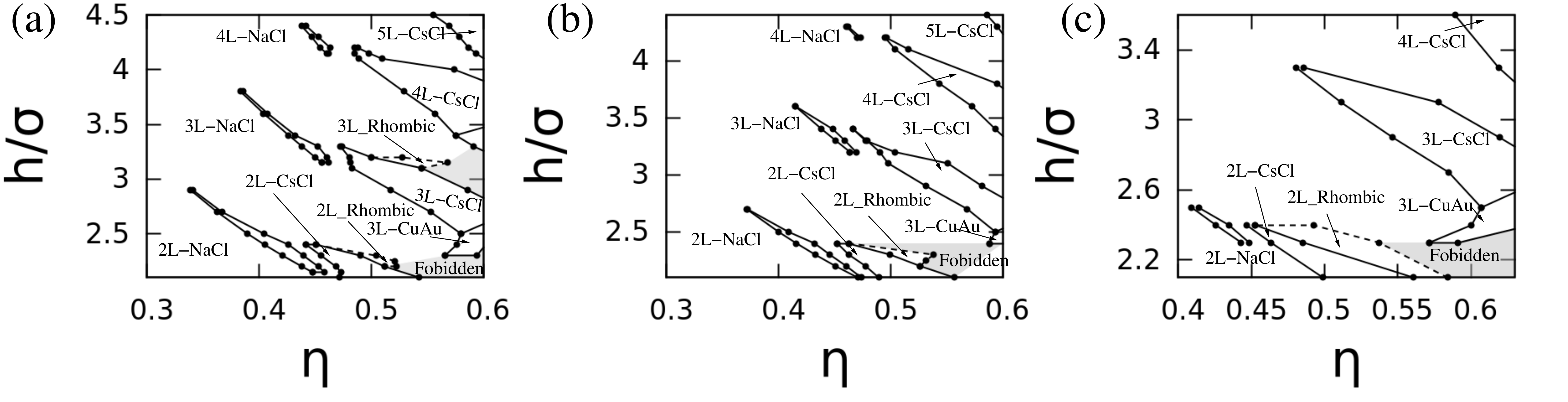}
\caption{Low temperature phase diagrams at $T^* = 0.1$ for confined oppositely charged colloids in the representation of $h/\sigma$ versus $\eta$ with various inversed screening length $\kappa \sigma = 3$  (a), 4 (b) and 6 (c). 
}\label{fig5} 
\end{figure*}

Next we calculate the low temperature phase diagrams at $T^* = 0.1$ for oppositely charged colloids confined between two parallel hard walls of different separation $h$ for various $\kappa$. As shown in Fig.~\ref{fig5}, at high density the stable phase always has the body-center-cubic (BCC), i.e., CsCl, face-center-cubic (FCC), i.e. CuAu, or multi-layer rhombic symmetry, which agrees with the phase behaviour of confined hard spheres at high pressure~\cite{fortini2006phase}. When decreasing the density of system, different low density phases appear in the phase diagrams in Fig.~\ref{fig5}. One can see that when $\kappa = 3 \sigma^{-1}$, with increasing $h/\sigma$ from about 2 to 4.5, the low density crystal phase always has the simple cubic symmetry, and the number of layers depends on the confinement height $h$, in which the maximal layers of stale simple cubic (NaCl) crystals at $T^* = 0.1$ and $\kappa = 3\sigma^{-1}$ is four. As shown in Fig.~\ref{fig4}b and c, the maximal layers of simple cubic crystals decreases with increasing $\kappa$, which agrees qualitatively with the confined system at ground state (Fig.~\ref{fig2}d). However, because of the effect of vibrational entropy, the maximal layers of stable simple cubic (NaCl) crystal at finite low temperature could be larger than that at the ground state.

\begin{figure*}[t!]
\centering
\includegraphics[width=\textwidth]{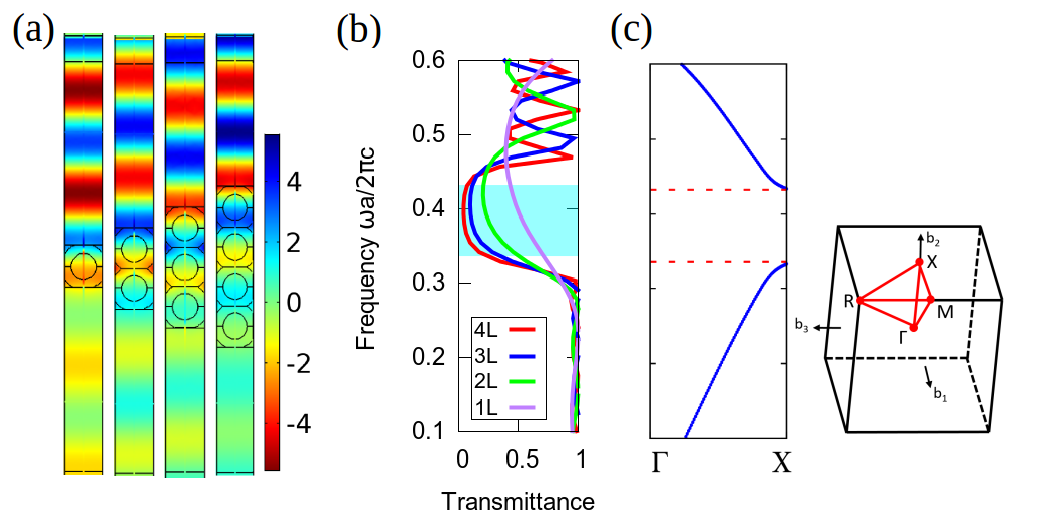}
\caption{
(a) Side view of optical simulation of normal incident lights with the frequency $\omega a/2\pi c=0.392$ upon an inverse 1-, 2-, 3-, and 4-layer simple cubic crystal of $a/\sigma = 5/6$ with $\sigma$ and $a$ the size of spherical bubble and the lattice constant of the crystal, respectively. The source incident light propagates from top to bottom, and the dielectric contrast for the inverse simple crystal is $\epsilon_r = 6$ with $c$ the speed of light. The color represents the intensity of the magnetic field (A/m).
(b) Transmission spectra of inverse 4-layer (red), 3-layer (blue), 2-Layers (green) and mono-layer (purple) simple cubic crystals planar lights at $a/\sigma = 5/6$ and $\epsilon_r=6.0$.
(c) The band structure along $\Gamma-\mathrm{X}$ direction, i.e., the direction perpendicular to the square plane, in the inverse simple cubic crystals with $a/\sigma = 5/6$ and $\epsilon_r=6.0$. Inset: the Brillouin zone of the simple cubic crystal~\cite{SETYAWAN2010299}.
}\label{fig6}
\end{figure*}

\begin{figure}[htbp]
\centering
\includegraphics[width=0.5\textwidth]{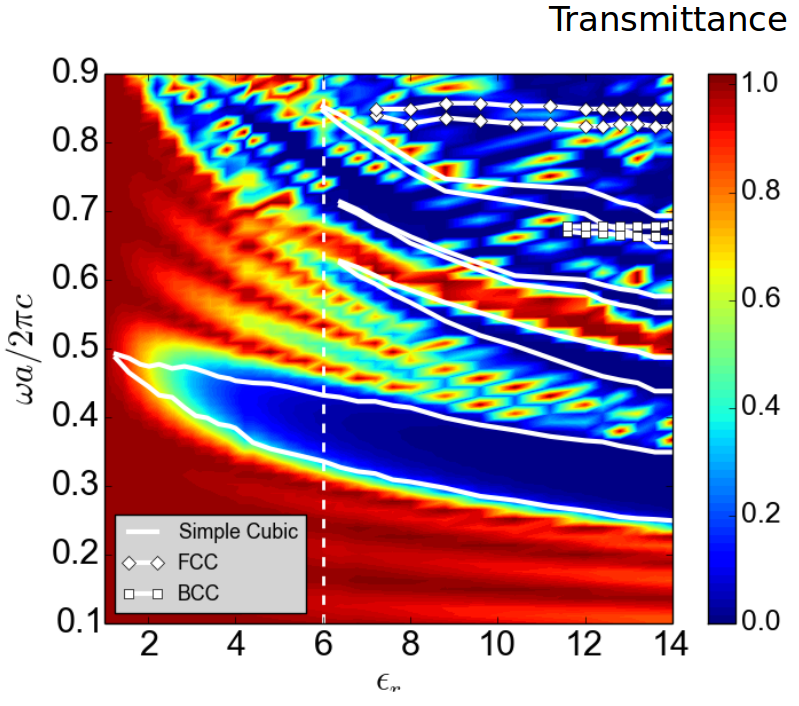}
\caption{Heat map of transmittance for the 4-layer simple cubic crystal in the representation of $\omega a/2\pi c$ versus $\epsilon_r$ at $a/\sigma = 5/6$. 
Solid lines enclose the photonic bandgap regions of the bulk simple cubic, FCC and BCC crystals in the direction perpendicular to the confining walls, i.e., the shadow planes in Fig.~\ref{fig1}b-d.
}\label{fig7} 
\end{figure}

\subsection{Photonic bandgaps of multi-layer simple cubic crystals}
Furthermore, we investigate the photonic property of obtained multi-layer simple cubic crystals. We first construct an inverse multi-layer simple cubic crystal, in which the spherical air bubbles of diameter $\sigma$ form a multi-layer simple cubic crystal with the lattice constant $a/\sigma = 5/6$. Experimentally, these inverse crystals can be obtained by selectively removing the colloidal particles after infiltrating the space between colloids using materials with the dielectric contrast $\epsilon_r$, in which the colloidal spheres are slightly overlapped, by sintering the particles in the crystal~\cite{blanconature2000}. We perform optical simulations with a normal incident light source on one side of the confined crystal. One can obtain the distribution of magnetic field $\mathbf{H}(\mathbf{r},t) = \mathbf{H}(\mathbf{r})e^{i\omega t}$ by numerically solving the simplified Maxwell equation in the frequency domain with the Floquet periodic boundary conditions in $x-y$ directions~\cite{joannopoulos2011p}
\begin{equation}\label{maxeq}
\nabla \times \left[ \frac{1}{\epsilon(\mathbf{r})} \nabla \times \mathbf{H}(\mathbf{r}) \right] = \left( \frac{\omega}{c}\right)^2 \mathbf{H}(\mathbf{r}),
\end{equation}
where $c$ and $\omega$ are the speed and frequency of light, respectively. The resulting $\mathbf{H}(\mathbf{r})$ for inverse multi-layer simple cubic crystals with $a/\sigma = 5/6$, $\omega a/2\pi c=0.392$, and $\epsilon_r = 6$ is shown in Fig.~\ref{fig6}a. With this, we calculate the transmittance of light with different frequency through the crystal by constructing perfectly matched layers~\cite{BERENGER1994185} on the top and bottom of the simulation box to absorb the excess reflected and transmitted light. The obtained transmission spectra of multi-layer inverse simple cubic crystal with $\epsilon_r = 6$ is shown in Fig.~\ref{fig6}b. Compared with the corresponding band structure of the bulk simple cubic crystal in $\Gamma-\mathrm{X}$ direction (Fig.~\ref{fig6}c), i.e. the direction perpendicular to the confining plane ~\cite{lei2018self}, with increasing the number of layers, the transmittance forbidden region approaches the band gap in the bulk simple cubic crystal, and the 4-layer simple cubic crystal can already almost reproduce the photonic band gap in the bulk crystal. In Fig.~\ref{fig7}, we plot the heat map of the transmittance for 4-layer simple cubic crystal in the representation of $\omega a/2\pi c$ versus $\epsilon_r$. Compared with the band gap obtained in the bulk simple cubic crystal in $\Gamma-\mathrm{X}$ direction, one can see that 
the transmittance forbidden region for the 4-layer simple cubic crystal persists to the low dielectric contrast $\epsilon_r \simeq 4$ close to the photonic band gap of the corresponding bulk crystal. 
For comparison, we also calculate the band gaps for bulk CsCl (BCC) and CuAu (FCC) crystals in the direction perpendicular to the confinement in Fig.~\ref{fig6}c.
Compared to the band gaps of FCC and BCC crystals, the band gaps for multi-layer simple cubic crystals are much larger and open at low frequency range, i.e., $0.2 \le \omega a/2\pi c \le 0.5$, suggesting that they are robust against disorder~\cite{zhangprb2000}.

\section{Conclusions}
To conclude, by using computer simulation, we have investigated the self-assembly of oppositely charged colloids confined in two parallel hard walls, in which we found stable multi-layer simple cubic crystals. With decreasing the inverse screening length $\kappa$, the number of layers of stable confined simple cubic crystals increases, and with $\kappa = 3\sigma^{-1}$, at low temperature, one can stabilize upto 4-layer simple cubic crystal, of which the inverse structure has large photonic band gaps open at low frequency with the dielectric contrast as low as $\epsilon_r \simeq 4$. This suggests new possibilities of fabricating photonic devices using low dielectric contrast materials like TiO$_2$ ($\epsilon_r \simeq 6.8$), SiC ($\epsilon_r \simeq 7.3$), ZnS  ($\epsilon_r \simeq 5.6$), GaN ($\epsilon_r \simeq 5.8$), ZnO ($\epsilon_r \simeq 5.8$), silica ($\epsilon_r \simeq 3.9$), etc., which were difficult for conventional colloidal self-assembly methods. {Therefore, the kinetic pathway of phase transitions, i.e. nucleation and melting, of these confined open crystals may be interesting for future investigation.}


\section{acknowledgments}
\begin{acknowledgments}
This work is supported by Nanyang Technological University Start-Up Grant (NTU-SUG: M4081781.120), the Academic Research Fund from Singapore Ministry of Education (M4011616.120 and M4011873.120), and the Advanced Manufacturing and Engineering Young Individual Research Grant (A1784C0018) by the Science and Engineering Research Council of Agency for Science, Technology and Research Singapore.
\end{acknowledgments}

\bibliographystyle{ACS}
\bibliography{reference_charged}

\end{document}